\documentstyle[amsfonts,aps,prl,twocolumn,epsf]{revtex}

\newcommand{\beq}{\begin{equation}}
\newcommand{\eeq}{\end{equation}}
\newcommand{\beqa}{\begin{eqnarray}}
\newcommand{\eeqa}{\end{eqnarray}}

\begin{document}

\title{Energy distribution of maxima and minima in a one-dimensional
random system}

\author{
Andrea Cavagna\thanks{E-mail: a.cavagna1@physics.ox.ac.uk},
Juan P. Garrahan\thanks{E-mail: j.garrahan1@physics.ox.ac.uk. On leave
from Department of Physics, University of Bue\-nos Air\-es, Arg\-en\-tina}
and Irene Giardina\thanks{E-mail: i.giardina1@physics.ox.ac.uk}
}

\address{
Theoretical Physics, University of Oxford,
1 Keble Road, Oxford, OX1 3NP, UK
\vskip 0.5 truecm
}

\address{\rm
We study the energy distribution of maxima and minima of a simple
one-dimensional disordered Hamiltonian.  We find that in systems with
short range correlated disorder there is energy separation between maxima 
and minima, such that at fixed energy only one kind of stationary points 
is dominant in number over the other. On the other hand, in the case of 
systems with long range correlated disorder maxima and minima are 
completely mixed.
\vskip 0.3 truecm}

\date{\today}

\maketitle

When the statistical properties of a system are stu\-died, great
attention is usually devoted to its ground state and to the first
excited states. Moreover, in the case of random systems it is known
that, in addition to the lo\-west energy states, also metastable states
are important, especially for the dynamical evolution of the system.
As a consequence, the physical understanding of disordered models is
most of the time founded on the properties of absolute as well as local
{\it minima} of the Hamiltonian, while the role of stationary points of
different nature is in general disregarded.  In spite of this, the
conviction is growing up that stationary points different from minima
do have an importance, both from a dynamical and a static point of view.

Many different disordered systems display an off-equilibrium dynamical
behavior which is suitable to be interpreted in terms of non-trivial
structure of their phase space.  Among these we find structural glasses
\cite{vetri}, spin glasses \cite{spin,francesi}, 
random manifolds \cite{manifolds} and neural networks \cite{neural}.  
In all these cases the geometric structure of the energy 
landscape is often invoked in order to give at least an 
intuitive picture of the relaxational dynamics.  
In this context it is clear that also unstable stationary 
points must be taken into consideration.  
For instance, the pre\-sence of flat directions in the phase space, 
marking a borderline between the last stable minima and the 
first unstable saddles, has been proposed as a possible 
explanation of slowness in glassy systems 
\cite{laloux,noiselle,sciortino}.

In the light of these considerations, we believe it 
is important to understand the connections between the 
physical properties of a disordered system and the 
geometrical structure of {\it all} the stationary 
points of its Hamiltonian.
Unfortunately, it is in general very difficult in an 
$N$-dimensional model to discriminate the stationary 
points according to their degree of instability.
In this Letter we will thus focus on a simple one-dimensional 
case and exactly compute the average energy distributions of 
maxima and minima. 
We will find a simple connection between the nature of the 
disorder which rules the physics of the system and the mutual 
distributions of the stationary points.
Despite its simplicity, we expect the model studied here 
to capture at least some of the main features of more
general problems.

Let us consider the one-dimensional random Hamiltonian 
\cite{villain,toy}, 
\[
H(x)=\frac{1}{2} m x^2 + V(x) \ ,
\]
where the position $x$ is a real variable and the mass $m$ is a
parameter. $V(x)$ is a Gaussian random potential, with zero average and
variance $\overline{V(x_1)V(x_2)}= G(x_1-x_2)$, with $G(x)=G(-x)$.  
The statics and the dynamics of this model have been studied both for
the one-dimensional case \cite{toy} and for the more general
$N$-dimensional case \cite{mepapot,ledu,franzpot}.  

The number of stationary points of $H$ is determined by the competition
between the random potential and the harmonic mass term. This number is
large for small $m$, whereas only one single mi\-ni\-mum is present at
large $m$.  The physical properties of this model are encoded in the
function $G$. In order to understand its meaning we consider the
average displacement $\left[ \Delta(d)\right]^2 = \overline{\left[
V(x_1)-V(x_2) \right]^2} = 2 G(0) - 2 G(d)$, where $d=(x_1-x_2)$ is the
distance.  Once introduced $\Delta$ it is natural to define two
different classes of random potentials.  If  $\Delta(d)$ goes to a
finite value $\Delta(\infty)$ for $d\to\infty$, then the memory is lost
after a finite distance and $V$ is called {\it short range} (SR).  On
the other hand, if $\Delta(d)\sim d^\gamma$ ($\gamma >0$), then the
displacement grows indefinitely with $d$ and the potential is {\it long
range} (LR).  In the SR case we can assume without loss of generality
that $G(x)$ is a positive even function which is zero at in\-fi\-ni\-ty, so
that $\Delta(\infty)=\sqrt{2G(0)}$.  In the LR case we have to be more
careful, since a diverging displacement would require $G(d)\to -\infty$
for $d\to \infty$, which is incompatible with the condition of having a
positive kernel in the functional distribution of $V$.  In order to
correctly define the LR model we must put the system in a box of size
$L$ and define $G_L(x)$ through its Fourier transform,
\beq
G_L(x)=\frac{1}{\pi} \int_{1/L}^\infty dq \,\hat G(q) \, e^{iqx} \ .
	\label{defa}  
\eeq
The function $\hat G(q)$ must be positive and for the LR case must be
{\it not} integrable in zero.  In order to avoid any ultraviolet
divergence we can assume both for the SR and LR cases $\hat G(q)$ to
decay at infinity faster than any power.  
We can thus define these two classes of models simply in terms of 
the behavior of $\hat G$ at zero momentum.  
In this way the LR case is well defined: $\Delta_L(d)$ increases 
indefinitely with $d$ and $\overline{V_L(x)^2}=G_L(0)$ diverges with $L$, 
as expected, since in a LR random potential the uncertainty on
the height $V$ of one single point $x$ increases with the size $L$ 
of the system, while it remains finite in a SR potential.
Our analysis will not depend on the explicit form of $\hat G$.

Let us denote by ${\cal N}^V_k(E,m)\, dE$, the number of 
sta\-tio\-na\-ry
points of $H(x)$ with degree of instability equal to $k$ ($k=0$ for
minima, $k=1$ for maxima), which have energy between $E$ and $E+dE$,
for a given mass $m$.  The superscript $V$ indicates that this 
distribution
corresponds to the sample $V$. Eventually we shall average over $V$.
The distribution ${\cal N}^V_k(E,m)$ is given by,
\[
{\cal N}^V_k(E,m) = \int dx \,\delta(H') \, \delta(H-E)  
	\, |H''|  \, \delta\left[\theta(-H'')-k\right] \; .
\]
In order to handle the modulus and the $\theta$-function we 
use the following relations, 
\beqa
\theta(-H'') &=& \frac{1}{2\pi i}\, \lim_{\epsilon\rightarrow 0}\,
	\left[ \log(H''-i\epsilon)-\log(H''+i\epsilon) \right] \; ,
	\nonumber \\
|H''| &=& \lim_{\epsilon\rightarrow 0} \,
	(H''+i\epsilon)^\frac{1}{2} (H''-i\epsilon)^\frac{1}{2} \; .
	\label{rela}
\eeqa
Using an integral representation for the $\delta$-function, we can  
write,
\beqa
\lefteqn{
	I \equiv |H''|  \,\delta\left[\theta(-H'')-k\right] = }
	\nonumber \\ 
&&	\int \,d\mu \,e^{ik\mu} \,
	  (H''+i\epsilon)^{\frac{1}{2}+\frac{\mu}{2\pi}}
	\,(H''-i\epsilon)^{\frac{1}{2}-\frac{\mu}{2\pi}} .
	\nonumber
\eeqa
The last two factors can be rewritten using the identity,
\[	(H''\pm i\epsilon)^{n_\pm} = 
	\int d\bar\chi^b_\pm \,d\chi^b_\pm \,
	\exp\left(-\sum_{b=1}^{n_\pm} 
	\bar\chi^b_\pm (H''\pm i\epsilon) \chi^b_\pm\right) , 
\]
where $\bar\chi^b_\pm$ and $\chi^b_\pm$ are Grassmann variables and the 
analytic continuation $n_\pm \to (1/2 \pm \mu/2\pi)$ must be done.
As a next step we define the Grassmann vector \cite{jorgesolo}: 
$\psi_a \equiv (\chi^1_+\dots \chi^{n_+}_+,\chi^1_-\dots \chi^{n_-}_- )$,
which allows us to write, 
\[
	I = \int \,d\mu \,e^{ik\mu} \int d\bar\psi_a \,d\psi_a \,
	\exp\left(-\sum_{a=1}^n 
	\bar\psi_a(H''+i\epsilon_a)\psi_a\right) ,
\]
where the  vector $\epsilon_a$ is split into two parts: 
$\epsilon_a=\epsilon$ for $a\le n_+$, 
$\epsilon_a=-\epsilon$ for $a>n_+$, and $n = (n_+ + n_-) \to 1$. 
Note that this replica approach can be easily generalized to 
$N$ dimensions.

Let us introduce in the expression for
${\cal N}_k^V$ the Lagrange multipliers $\lambda$ and $\omega$, 
to represent respectively $\delta(H')$ and $\delta(H-E)$ . 
The $V$-dependent part then becomes: 
$\exp[(i\omega+i\lambda\partial_x-\bar\psi_a\psi_a
\partial_x\partial_x)V(x)]$, which can be averaged over the 
Gaussian distribution of $V$. 
This produces a quartic term $(\sum_a\bar\psi_a\psi_a)^2$, 
that can be made quadratic by means of a Hubbard-Stratonovich 
transformation, introducing an auxiliary variable $y$.
It is now possible to perform all the Gaussian integrals
over $(\lambda,x,\psi)$. This gives a term 
$(m+y+i\epsilon)^{1/2+\mu/2\pi}(m+y-i\epsilon)^{1/2-\mu/2\pi}$, 
which, using back relations (\ref{rela}), can be written as  
$|m+y|\,\exp[-i\mu\,\theta(-m-y)]$. Integrating over $\mu$ 
we finally obtain the average distributions 
${\cal N}_k(E,m)\equiv\overline{{\cal N}_k^V(E,m)}$,
\beqa
&&	{\cal N}_0(E,m) = \int_{-m}^{+\infty}  dy \, F(y,E,m) \ , 
	\label{numeri0} \\
&&	{\cal N}_1(E,m) = \int_{-\infty}^{-m}  dy  \,F(y,E,m) \ , 
	\label{numeri1}
\eeqa
with,
\beqa
&&	F(y,E,m)=\frac{\left| m+y \right|}{\sqrt{m}}\, 
	\frac{e^{-\frac{y^2}{2 a_2}}}{\sqrt{2\pi a_2}} \ \times   
	\nonumber \\
&&	\times \,\int \frac{d\omega}{2\pi} 
	\,\frac{e^{  -\frac{1}{2} 
	\left( a_0 - a_1^2/a_2 \right) \omega^2 
	\,+\, i\omega E + i \omega y\, a_1/a_2 } }
	{\sqrt{m+ia_1\omega}} \ ,
	\label{megaultra}
\eeqa
where $a_0=G(0)$, $a_1=-G''(0)$ and $a_2=G''''(0)$. 
The difference ${\cal D}(E,m)= {\cal N}_0(E,m)-{\cal N}_1(E,m)$ 
between mi\-ni\-ma and maxima has a much simpler expression,
\[
{\cal D}(E,m)=
	\frac{1}{\sqrt{m}}
	\int \frac{d\omega}{2\pi} \, \sqrt{m+ia_1\omega} 
	\,e^{-\frac{1}{2}a_0\omega^2 + i\omega E} \ .
\]
By integrating equations (\ref{numeri0}) and (\ref{numeri1})
over the energy we get the
total number of minima and maxima, ${\cal N}_0(m)$ and 
${\cal N}_1(m)$, at a given value of the mass. 
Note that, as required by the Morse theorem \cite{morse},
the {\it total} number of minima minus maxima is equal to one,
that is ${\cal N}_0(m)-{\cal N}_1(m)=\int dE \; 
{\cal D}(E,m)= 1$.  
The explicit expression for the total number of minima is,
\[
{\cal N}_0(m) =
	\frac{1}{2}+ \frac{1}{2} {\rm erf} 
	\left( \frac{m}{\sqrt{2 \, a_2}} \right)
	+ \frac{1}{m} \sqrt{\frac{a_2}{2 \pi}} 
	\exp \left( - \frac{m^2}{2\, a_2} \right) \ .
\]
${\cal N}_0(m)$ is a smooth function of $m$ which goes to one when $m$
goes to infinity and starts increasing very steeply at masses smaller
than $m \sim \sqrt{a_2}$.  This value of the mass marks a crossover
from the region where only one minimum exists to the region where many
different minima (and maxima) appear.  As expected this mass is the
same critical mass as the $N$-dimensional
 mean-field case, $m_c=\sqrt{a_2/3}$, 
where a glassy transition occurs \cite{toy,ledu}.  In
the following we will always consider $m < m_c$.

\begin{figure}
\begin{center}
\leavevmode
\epsfxsize=3in
\epsffile{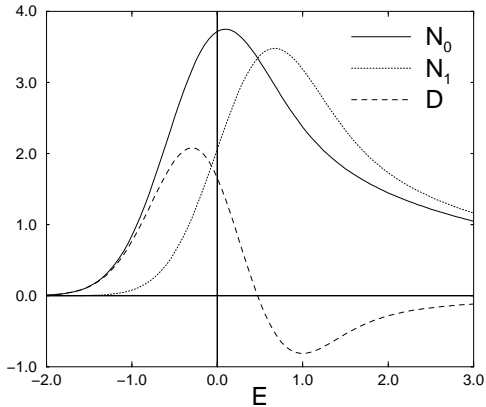}
\caption{SR potential: ${\cal N}_0(E,m)$, ${\cal N}_1(E,m)$ 
and ${\cal D}(E,m)$, functions of the energy $E$, at $m<m_c$.
Here $\hat G(q)=e^{-q}$.}
\label{fig1}
\end{center}
\end{figure}

We analyze now our results, starting with the SR potential.  
In Fig.\ref{fig1} we plot ${\cal N}_0(E,m)$, ${\cal N}_1(E,m)$ and  ${\cal
D}(E,m)$ functions of the energy $E$, for $m<m_c$.  
\noindent
The first thing we
notice is that the two curves of mi\-ni\-ma and maxima are quite separated
one with respect to the other, so that their peaks do not overlap.  As a
consequence ${\cal D}(E,m)$ gives by itself a rather clear picture of
the distribution of the different stationary points and at low energies it
approximates well  ${\cal N}_0(E,m)$.  This is important because ${\cal
D}$ is always very simple to compute, being ${\cal D}(E,m)=\int dx \,
\delta(H')\, H''\, \delta(H-E)$. Thus, the computation of this quantity
does not require the modulus, nor the $\theta$-function, which are in
general very difficult to treat.  In other words, in the SR case there
is a partial decoupling between maxima and minima, which is sharper
the lower is the energy. As a consequence, at fixed
energy only one kind of stationary points is dominant over the
other and ${\cal N}_0(E,m)\sim {\cal D}(E,m)$, for low enough $E$.  
It is remarkable that this holds for the SR potential.
Indeed, it has been proved in \cite{noiselle} that in
the $N$-dimensional $p$-spin spherical spin glass, which 
belongs to the SR class \cite{ledu}, an identical phenomenon occurs:  
in that mean-field model, besides minima and maxima there are saddles of 
any order, but at fixed energy only one kind of stationary points 
is dominant over the other, so that the number of minima of the mean-field 
free energy (and therefore of states) can be safely
calculated via the approximation ${\cal N}_0(E)\sim {\cal D}(E)$.
We note that this same approximation has been used many times in the 
context of spin glasses, regardless of its grounding \cite{kurchan}.

Let us now turn to the LR potential. We stress that the model is
defined as long range (and thus it is different from the SR case) only
in the limit $L\to\infty$. From Eq.(\ref{defa}) we know that  
$a_0=G_L(0)=\overline{V(x)^2}$ diverges with $L$.
As mentioned above, the physical meaning of this is that the uncertainty
on the value of $H$ in any point $x$ diverges when $L$ goes to infinity.
As a consequence, the energy $E$ is no longer a good variable to label the
height of the sta\-tio\-na\-ry points. In order to keep everything well defined
in the limit $L\to\infty$ it is thus necessary to measure the energy 
in units of the natural diverging scale,
that is $\sqrt{a_0}$. Therefore, we must define a rescaled energy 
${\cal E}=E/\sqrt{a_0}$ and study the distributions of maxima and 
minima as functions of $\cal E$. Denoting these new distributions 
by $P_0$ and $P_1$, we have, 
$P_k({\cal E},m)\, d{\cal E} \equiv {\cal N}_k(E,m) \, dE$. 
We stress that $\cal E$ is the only variable we can 
sensibly regard as the energy for the LR potential.
Note, on the other hand, that this rescaling is irrelevant 
for the SR case, where $a_0$ is finite.
Taking the limit $L\to\infty$ in Eq.(\ref{megaultra}) we find,
\beq
P_{0,1}({\cal E},m) \to \ {\cal N}_{0,1}(m) \
	\frac{1}{\sqrt{2\pi}} \ e^{-\frac{1}{2} {\cal E}^2} \ .
	\label{minkia}
\eeq
This equation shows that in the LR case the two distributions are
just the {\it same} function, scaled by the total number ${\cal N}_0(m)$
or ${\cal N}_1(m)$. Maxima and minima are no longer separated 
in energy. Indeed, for $m\ll m_c$, we have ${\cal N}_0(m)/{\cal N}_1(m) 
\sim 1$ and the two curves collapse one on to the other.
The conclusion is that when the total number of stationary points 
is large in a LR system, maxima and minima are completely mixed 
together, so that at each given energy they are equally numerous.
Thus, in stark contrast with the SR case, no decoupling of the stationary 
points occurs, no matter how low is the energy.

A further step is necessary to prove that this mixing 
in the LR case is a {\it typical} be\-ha\-vior and
not simply an artifact coming from the average. 
Indeed, it is possible to think of a system where 
sample by sample maxima and minima are well separated, 
but where the mixing described above appears only
after averaging over different samples. 
As an example, we consider the family of Hamiltonians
$H_w(x)=\sin(x)+w$, where $w$ is a random variable with zero 
average and variance $\sigma$. 
It is clear that for each sample maxima and mi\-ni\-ma are 
perfectly separated, since ${\cal N}^w_{0,1}(E)=\delta(E\pm 1-w)$. 
However, averaging over $w$ we obtain two distributions with 
separation between their averages equal to 2 and variance $\sigma$. 
Thus, if we rescale the energy by a factor $\sqrt{\sigma}$ and take the 
limit $\sigma\to\infty$, we would conclude that there is 
mixing between maxima and minima, which is sample by sample false.

In order to prove that the LR potential does not correspond to such 
an artifact, we consider the statistics of the extreme values of the 
Hamiltonian.
Let us  define the two distributions 
$A_0(E)\equiv\overline{\delta(E-E_{MIN})}$
and $A_1(E)\equiv\overline{\delta(E-E_{MAX})}$, 
where $E_{MIN}$ and $E_{MAX}$ are the energies 
of the absolute minimum and maximum of $H(x)$
(we consider as absolute maximum the highest local maximum).
The separation between these two distributions is $\Delta A=
\overline{E_{MAX}-E_{MIN}}=
\langle E \rangle_{A_1}-\langle E \rangle_{A_0}$,
and let $S$ be their variance. 
Consider now the ratio $\Delta A/\sqrt{S}$.
It is easy to see that in the artificial case described 
above this ratio goes to zero when the variance of the disorder
$\sigma$ goes to infinity, since  $\Delta A$ is finite, 
whereas $S$ diverges as $\sigma$.
On the other hand, for the LR potential the ratio  
$\Delta A/\sqrt{S}$ remains finite in the limit $L\to \infty$.
This is simple to prove exactly for $\gamma=1/2$, which corresponds 
to a Brownian random potential of size $L$ \cite{villain}. For the 
general case the idea is that in a LR potential both  
$\Delta A$ and $\sqrt{S}$ diverge as  $L^\gamma$ \cite{book},
as can also be checked by means of numerical simulations.  
The divergence of $\Delta A$  implies that the variance of the 
energy distribution of maxima and minima is {\it sample by sample} 
diverging as well with $L$. Therefore, unlike what happens in the 
case of the artifact, the average scenario of the long range potential, 
where maxima and minima are completely mixed in energy, 
{\it is} the typical one.

In this Letter we have shown that the energy distribution of 
maxima and minima in a one-dimensional random system is radically 
different whether the disorder is long range or short range 
correlated.
An important issue is the extension of this investigation to
$N$ dimensions, where we expect to find a qualitative similar 
behavior, at least at the mean-field level.

Indeed, as mentioned above, the $p$-spin spherical 
spin glass is a clear example of an $N$-dimensional 
mean-field SR system where at sufficiently low 
energies a decoupling between stationary points of 
different nature does occur.
Moreover, a crucial feature of this SR model is that the 
a\-symp\-to\-tic dynamical energy reached by the system  
is exactly the same energy where 
the stationary points decouple \cite{noiselle}, and
is larger than the equilibrium one, so that the
dynamics never gets to the equilibrium landscape.

On the other hand, let us assume that for a LR $N$-dimensional 
model a generalization of equation (\ref{minkia}) is valid,
so that {\it all} the sta\-tio\-na\-ry points 
collapse on one single distribution, since saddles 
of any degree of instability will be bounded in energy by 
maxima and minima.
This allows us to put forward the following hypothesis: 
if, as indicated by our results, in a LR system 
there is no separation at all between different stationary 
points, then no decoupling energy can exist, which means  
no energy level capable of trapping the system.
In such a situation we would expect the dynamics to reach the minimum 
available energy, that is the equilibrium energy. 

An evidence of this conjecture can be found in the context of
mean-field models for spin glasses. 
Here two very different classes of systems exist, a first class 
where the dynamical energy is larger than the equilibrium 
one and a second class where these two energies are the same.
What has been noted in \cite{ledu}, is that the first class corresponds 
to systems with SR disorder (as the $p$-spin model), 
whereas to the second class belong LR disordered systems. 
This correspondence finds its natural explanation in the framework 
we have depicted above: the inability of a SR system to dynamically 
reach its equilibrium energy is due to the existence of an 
energy level below which stationary points of different nature are
decoupled, while this cannot happen in the LR case.
As we have tried to show in this Letter, whether this decoupling 
occurs or not, is an information encoded in the energy distribution of 
{\it all} the stationary points.

\acknowledgements
We thank J. Kurchan, G. Parisi and D. Sherrington 
for very important observations and suggestions, 
and L.F. Cugliandolo for reading the manuscript and for
many useful comments.  We also thank A. Baldassarri, L.B. Ioffe,
L. Tremolada and P. Verrocchio for useful discussions.
A.C. and I.G. wish to thank for the kind hospitality the Department of
Physics of the ENS of Lyon, where part of this work was done.  The work
of A.C. and I.G. was supported by EPSRC Grant GR/K97783.  The work of
J.P.G. was supported by EC Grant ARG/B7-3011/94/27.



\begin{thebibliography}{}

\bibitem{vetri} 
C.A. Angell,
Science {\bf 627}, 1924 (1995);
F.H. Stillinger, 
Science {\bf 267}, 1935 (1995);
S. Sastry, P.G. Debenedetti and F.H. Stillinger, 
Nature {\bf 393}, 554 (1998).

\bibitem{spin} 
M. M\'ezard, G. Parisi and M. A. Virasoro,
{\em Spin Glass Theory And Beyond}
(World Scientific, Singapore, 1986).  

\bibitem{francesi}  
J.-P. Bouchaud, L.F. Cugliandolo, J. Kurchan and M. M\'ezard, 
in {\it Spin-glasses and random fields}, A.P. Young Ed. 
(World Scientific, Singapore, 1997).

\bibitem{manifolds} G. Blatter, M.V. Feigel'man, V.B. Geshkenbein,
A.I. Larkin and V.M. Vinokur,
Rev. Mod. Phys. {\bf 66}, 1125 (1994).

\bibitem{neural} V. Dotsenko, 
{\it An Introduction to the Theory of Spin Glasses
and Neural Networks} 
(World Scientific, Singapore, 1995).

\bibitem{laloux} 
J. Kurchan and C. Laluox, 
J. Phys. A {\bf 29}, 1929 (1996).

\bibitem{noiselle}
A. Cavagna, I. Giardina and G. Parisi, 
Phys. Rev. B {\bf 57}, 11251 (1998).

\bibitem{sciortino} 
F. Sciortino, S. Sastry, P. Tartaglia, cond-mat/9805040.

\bibitem{villain} 
J. Villain, B. Semeria, F. Lan\c con and L. Billard,
J. Phys. C {\bf 16}, 6153 (1983).

\bibitem{toy} 
M. M\'ezard  and G. Parisi,
J. Phys. I France {\bf 2},  2231 (1992);
A. Engels,
Nucl. Phys. B {\bf 410} [FS], 617 (1993).

\bibitem{mepapot} 
M. M\'ezard and G.Parisi,
J. Phys. A {\bf 23}, L1229 (1990);  
J. Phys. I France {\bf 1}, 809 (1991).

\bibitem{ledu} L.F. Cugliandolo, J. Kurchan and P. Le Doussal,
Phys. Rev. Lett. {\bf 76}, 2390 (1996);
L.F. Cugliandolo and P. Le Doussal,
Phys. Rev. E {\bf 53}, 1525 (1996).

\bibitem{franzpot} 
S. Franz and M. M\'ezard,
Europhys. Lett. {\bf 26}, 209 (1994); 
Physica A {\bf 209}, 1 (1994).

\bibitem{morse}  
B. Doubrovine, S. Novikov  and A. Fomenko,  
{\it Geometrie Contemporaine} (Mir, Moscow, 1982).

\bibitem{jorgesolo} 
J. Kurchan, private communication (1998).

\bibitem{kurchan} 
For a discussion of this problem see J. Kurchan,
J. Phys. A {\bf 24}, 4969 (1991) and references therein.

\bibitem{book} G.R. Grimmett and D.R. Stirzaker, 
{\it Probability and Random Processes} (Clarendon Press, Oxford, 1982).

\end{thebibliography}
\end{document}